\documentstyle[epsfig]{mn}
\input{psfig}
\begin{document}
\def\ltsima{$\; \buildrel < \over \sim \;$}
\def\simlt{\lower.5ex\hbox{\ltsima}}
\def\gtsima{$\; \buildrel > \over \sim \;$}
\def\simgt{\lower.5ex\hbox{\gtsima}}

\title[The swan song: the disappearance of the nucleus of NGC~4051 and the
echo of the its past glory]
{The swan song: the disappearance of the nucleus of NGC~4051 and the
echo of its past glory}

\author[M. Guainazzi et al.]
{M. Guainazzi$^1$, F. Nicastro$^{2,3}$, F. Fiore$^{3,4}$, G. Matt$^5$, I. McHardy$^6$, A. Orr$^1$, P. Barr$^1$, \and A. Fruscione$^7$, I. Papadakis$^8$, A.N. Parmar$^1$, P. Uttley$^6$, G.C. Perola$^5$, L. Piro$^2$ \\ ~ \\
$^1$Astrophysics Division, Space Science Department of
ESA, ESTEC, Postbus 299, 2200 AG Noordwijk, The Netherlands\\
$^2$Istituto di Astrofisica Spaziale, C.N.R., Via Fosso del Cavaliere, I-00133 Roma, Italy \\
$^3$Osservatorio Astronomico di Roma, Via dell'Osservatorio 5, I-00040 Monteporzio Catone, Italy \\
$^4$BeppoSAX Science Data Center, Via Corcolle 19, I-00131 Roma, Italy \\
$^5$Dipartimento di Fisica ``E.Amaldi'', Universit\`a degli Studi Roma~3, Via della Vasca Navale 84, I-00146 Roma, Italy \\
$^6$Department of Physics \& Astronomy, University of Southampton, University Road, Southampton SO17 1BJ, UK \\
$^7$Harvard-Smithsonian Center for Astrophysics, 60 Garden Street, MS-70, Cambridge, MA 02138  USA \\
$^8$Skinakas Observatory, Physics Department, University of Crete, Herakelion, Crete, Greece \\
}

\maketitle
\begin{abstract}
On 9-11 May 1998,
BeppoSAX observed the low--luminous Seyfert 1 Galaxy NGC4051
in a ultra--dim X--ray state. The 2-10 keV flux (${\rm 1.26
\times 10^{-12}}$~erg~cm$^{-2}$~s$^{-1}$) was 
about 20 times fainter than its historical average value, and remained
steady along the whole observation ($\sim$2.3 days). The observed flat
spectrum ($\Gamma \simeq 0.8$) and intense iron line (EW$\simeq$600 eV)
are best explained assuming that the active nucleus has switched off,
leaving only a residual reflection component visible.
\end{abstract}


\section{Introduction}

X--ray spectra of Seyfert galaxies above a few keV are
dominated by two
components: a power law (hereinafter ``primary") component 
emitted very close to the black hole, 
and a secondary (hereinafter ``reflection") component arising 
from the reprocessing of the primary radiation by neutral,
optically thick matter (Pounds et al. 1990, Piro et al. 1990). The reflection 
component, produced by a Compton
scattering, forms a broad hump peaking at about 30 keV,
and an intense iron K$\alpha$ fluorescent line at 6.4 keV (Lightman
\& White 1988; George \& Fabian 1991; Matt, Perola \& Piro 1991). 
While in many Seyfert 1s at least part of the reprocessing occurs
in the inner accretion disc, as indicated by the observed relativistic 
effects
in the iron line profile (Tanaka et al. 1995; Nandra et al. 1997),
a further reprocessed component can arise if the nucleus is surrounded
by any large amount of distant, optically thick and neutral matter
(Ghisellini, Haardt \& Matt 1994), such as the ``torus''
expected in the unification scenario (Antonucci \& Miller 1985; Antonucci 
1993) or the dust lanes visible in the
high--resolution optical images of Seyfert galaxies (Malkan et al. 1998;
Maiolino et al. 1998).
While the reflection 
component from the accretion disc should lag the primary emission by
minutes or hours at most, the lag introduced by the distant reflector 
is probably of the order of at least weeks.
If the active nucleus suddenly
switches off, the latter component would then continue
echoing the primary component for some time, remaining for a while
the only witness of the active nucleus past activity. 
We have observed what could be precisely
this situation in the nearby (${\rm z=0.0023}$),
low--luminosity (${\rm < L_{2-10 keV} > \sim 5 \times 10^{41}}$~erg~s$^{-1}$)
Seyfert 1 galaxy NGC4051. BeppoSAX observed
the source on 9-11 May 1998
and measured a
constant flux along the whole observation
($\simeq 2.3$~days), which was a factor of five lower than the faintest
state ever observed (which lasted only a few thousands of
seconds, Uttley at al. 1998).
A contemporaneous and longer EUVE observation suggests that
such a faint state lasted for at least one week (Fruscione et al.,
in preparation).
The spectrum above
a few keV is completely dominated by a ``bare" reflection component
and implies a luminosity of the primary nuclear component
${\rm L_{nuc} < 0.14 \times 10^{41}}$~erg~s$^{-1}$.
This is the first detection of a so dramatic and long-lasting fading 
in luminosity from NGC4051, and provides one of
the first {\it direct} evidence for the
presence of large amount of circumnuclear cold, thick matter 
in Seyfert 1s environment.

\section{Data Reduction}

NGC4051 was observed by the Italian-Dutch satellite
BeppoSAX (Boella et al. 1997a) from 1998 May 9 09:56:44 UTC to May 11
18:05:06 UTC.
In this paper, data of the imaging instruments
(Low-Energy Concentrator Spectrometer,
LECS, Parmar et al. 1997; Medium-Energy Concentrator Spectrometer,
MECS, Boella et al. 1997), and of the collimated
Phoswitch Detector System (PDS, Frontera et al. 1997)
are presented.
Cleaned and linearized event files have been obtained with
the reduction software package {\sc Saxdas} (version 1.3.0), using
standard screening criteria (see {\it e.g.} Matt et al. 1997).
The PDS data have
been further screened with a temperature--dependent
rise--time threshold, which
allows a $\simeq 50\%$
reduction in instrumental background. The total net
exposure times are 55.9, 69.2 and 33.0~ks for the LECS, MECS and PDS,
respectively.

In the following,  uncertainties are quoted at 90\% level of confidence
for one interesting parameter ($\Delta \chi^2 = 2.71$);
energies are in the source rest frame; ${\rm H_0 =
50}$~km~s$^{-1}$~Mpc$^{-1}$ is assumed throughout.

\section{Spectral analysis}

LECS (MECS) spectra have been extracted in circular regions of
4 (2) arc minutes and rebinned in order to have at least 3 (2)
energy channels per resolution element. Background
spectra from blank sky fields and response matrices
publicly available at the Beppo SAX Science Data Center (SDC,
September 1997 release) have been employed throughout.
PDS spectra have been obtained by subtraction of the
``off-'' from the ``on-source'' intervals.
The net count rates are
$(1.12 \pm 0.05) \times 10^{-2}$~s$^{-1}$ in the LECS (0.1--4~keV),
$(1.09 \pm 0.04) \times 10^{-2}$~s$^{-1}$ in the MECS (1.8--10.5~keV)
and $0.123 \pm 0.027$~s$^{-1}$ in the PDS (13--50~keV).
No variability is observed both in the 0.1-3 and
3-10~keV energy bands,
contrary to the factor up to 50 typically observed
in various X--ray energy bands (McHardy et al. 1995;
Guainazzi et al. 1996; Uttley et al. 1998; Fruscione et al., in
preparation).

A fit of the 1.8--10.5~keV
MECS spectrum alone with a simple power--law model, absorbed
through a column of neutral Hydrogen with ${\rm N_H = N_{H_{Gal}} =
1.3 \times 10^{20}}$~cm$^{-2}$ (Elvis, Lockman \& Wilkes, 1989),
yields a very poor fit ($\chi^2 =99/36$~degrees of freedom, dof).
The main contribution to the residuals is due to a large emission
feature at ${\rm \simeq 6.4}$~keV. Adding a Gaussian profile to the
continuum significantly improves the quality of the fit
($\Delta \chi^2 = 46$). Best-fit parameters are: photon index
${\rm \Gamma = 0.78\pm^{0.37}_{0.13}}$, line centroid ${\rm E_c
= 6.57\pm^{0.16}_{0.13}}$~keV, line Gaussian dispersion ${\rm
\sigma=0.3\pm0.3}$~keV, line equivalent width ${\rm
EW = 1.9\pm^{0.9}_{0.5}}$~keV. The 2--10~keV flux is
$\simeq 1.4 \times 10^{-12}$~erg~cm$^{-2}$~s$^{-2}$.
Despite the approximate nature
of the spectral description, three points are
evident: a) BeppoSAX observed NGC~4051 in an ultra--dim
state, with a 2--10~keV flux
about a factor of 20 fainter than its historical average
(see Sect.~4);
b) the intermediate X--ray spectrum is much flatter than in any previously
reported observation (see {\it e.g.}: Fiore et al. 1992; Mihara et al. 1994,
Guainazzi et al. 1996) and in any other Seyfert 1 galaxy;
c) the iron line EW is several times higher than
typically observed in Seyfert 1s in their ``normal'' state
(Nandra \& Pounds 1994; Nandra et al. 1997) and in NGC4051 in
particular (Mihara et al. 1994; Guainazzi et al. 1996).

These pieces of evidence suggest that the
primary nuclear emission has switched off leaving 
the Compton reflection component (along with the iron fluorescent
line) as an echo of the past activity.
We therefore fitted the MECS (above 4~keV)
and PDS spectra with a ``bare'' reflection component only plus a Gaussian.
We used the {\sc Xspec} model {\verb!pexrav!}, assuming
an inclination of 0$^{\circ}$,
solar abundances, and no high--energy
cut--off in the intrinsic power--law; such assumptions do not
substantially affect the following results.
%-----------------------------Figure 3--------------------------------
\begin{figure}
\begin{center}
\epsfig{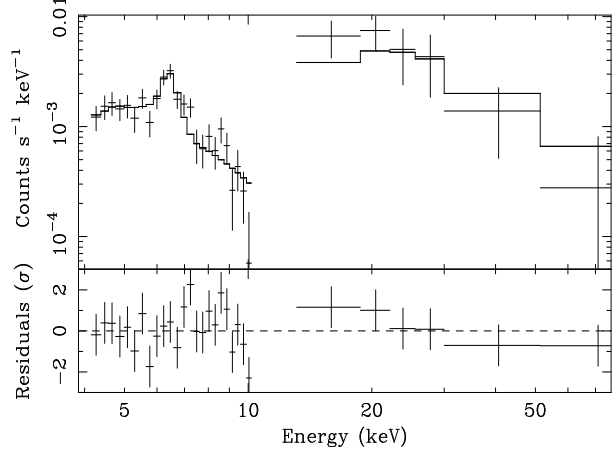}
\end{center}
\caption{MECS and PDS spectra ({\it upper panel})
and residuals in units of standard deviations ({\it lower
panel}) when a ``bare'' Compton reflection continuum model
+ Gaussian line model is applied.}
\label{fig3}
\end{figure}
%-----------------------------Figure 3--------------------------------

This model yields a significantly better fit
(${\rm \chi^2_{\nu} = 1.17}$) than the ones where
the continuum is fitted with a simple power-law (${\rm \chi^2_{\nu} = 1.38}$),
or with a strongly absorbed (${\rm N_H \sim 3 \times 10^{24}}$)
Seyfert-like (${\rm \Gamma \equiv 1.9}$) power-law + reflection
from a plane-parallel infinite slab (${\rm \chi^2_{\nu} = 1.36}$),
which both significantly underestimate the PDS counts
below 30~keV. Table~\ref{tab1} summarizes the best-fit results.
The spectral index of the illuminating (but now invisible) power--law
($\Gamma \simeq 1.92$, see Figure~\ref{fig3}) turns out to be
perfectly consistent with the typically observed value for 
NGC4051 (see Guainazzi et al. 1996 and references therein).
The 2--10~keV flux (luminosity) as inferred from the best-fit is
$(1.26 \pm 0.06) \times 10^{-12}$~erg~cm$^{-2}$~s$^{-1}$
([$2.84 \pm 0.14] \times 10^{40}$~erg~s$^{-1}$).
The 90\% upper limit on the 1~keV relative normalization
between the primary and reflected continua is $\simeq 2\%$.
This implies a nuclear flux decrease
of at least a factor of 35 in 1.5 years,
if compared with the average measured in the latest RXTE
long-look pointing (Uttley et al. 1998).
Alternatively, a column density of at least ${\rm 1.4 \times
10^{26}}$~cm$^{-2}$ is needed,
if the fading is due to absorption rather than a switch off
of the primary continuum.

The line centroid energy is consistent
with fluorescence from neutral or mildly ionized iron
(${\rm E_c = 6.50\pm^{0.12}_{0.10}}$~keV), with an
${\rm EW \simeq 600}$~eV. The width of the line is unconstrained, 
the 90\% upper limit on $\sigma$ being an inconclusive
320~eV. However, some residuals in the blue wing of the line
may suggest a line blending.
Adding a further narrow emission line
improves the quality of the fit only at a 93.7\% confidence level.
If we assume that an
iron ${\rm K_{\alpha}}$/${\rm K_{\beta}}$ pair is produced
in the same medium (and therefore with the same intrinsic width),
%-----------------------------Table 1--------------------------------
\begin{table*}
\begin{footnotesize}
\centering
\caption{
Best fit parameters for the May 1998 observation. Errors correspond to 
$\Delta\chi^2$=2.7. The iron line energy is given in the source rest frame.}
\label{tab1}
\vspace{0.05in}
\begin{tabular}{llcccccccc} \hline 
\# & Model & $\Gamma$ & ${\rm E_c^1}$ & ${\rm \sigma^1}$ & ${\rm EW^1}$ & ${\rm E_c^2}$ & ${\rm \sigma^2}$ & ${\rm EW^1}$ & $\chi^2/$dof \\
& & & (keV) & (eV) & (eV) & (keV) & (eV) & (eV) & \\ \hline
1 & PO & $0.76\pm^{0.17}_{0.15}$ &  &  &  & & & & 85.0/35 \\
2 & PO+BL & $0.75\pm^{0.21}_{0.17}$ & $6.52 \pm^{0.17}_{0.09}$ & $160\pm^{270}_{160}$ & $1400\pm500$ & & & & 44.1/32 \\ 
3 & PO+CR+BL & 1.9$^{\dag}$ & $6.50\pm^{0.09}_{1.46}$ & $< 6000$ & $600\pm^{300}_{200}$ & & & & 43.6/32 \\
4 & CR & $1.94\pm^{0.18}_{0.14}$ &  &  &  & & & & 56.8/35 \\
5 & CR+NL & $1.92\pm^{0.19}_{0.15}$ & $6.49\pm^{0.10}_{0..09}$ & 0$^{\dag}$ & 
$600\pm^{300}_{200}$ & & & & 38.4/33 \\
6 & CR+BL & $1.92\pm^{0.20}_{0.15}$ & $6.50\pm^{0.12}_{0.10}$ & 
$100\pm^{220}_{100}$ & $600\pm^{300}_{200}$ & & & & 37.4/32 \\ 
7 & CR+BL+BL & $1.92\pm^{0.18}_{0.09}$ & 6.4$^{\dag}$ & $< 150$ & $600\pm^{300}_{200}$ & 
7.1$^{\dag}$ & $\equiv \sigma^1$ & $400 \pm 300$ & 33.6/32 \\ \hline
\end{tabular}
\\
\noindent
$^{\dag}$~fixed \\
Notes: PO is a power-law,
CR is the ``bare'' Compton reflection continuum, NL is a narrow
Gaussian emission line, BL a Gaussian emission line where the
intrinsic width is left as a free parameter. Photoelectric
absorption with ${\rm N_H = N_{H_{Gal}} \equiv 1.3 \times 10^{20}}$~cm$^{-2}$
has been added to all the models above, except for the model \#~3, for which
${\rm N_H = (2.9 \pm 0.9) \times 10^{24}}$~cm$^{-2}$. In model \#~3 the
relative normalization between the primary and Compton reflected component
has been held fixed to 1.
\end{footnotesize}
\end{table*}
%-----------------------------Table 1--------------------------------
the best fit ratio of the
${\rm K_{\beta}}$/${\rm K_{\alpha}}$ intensities is $0.5\pm^{0.2}_{0.4}$,
higher than, albeit not formally inconsistent with, the theoretically
expected value ($\simeq$~11\%). Interestingly enough, the line flux
(${\rm I \simeq 1.0 \times 10^{-5}}$~photons~cm$^{-2}$~s$^{-1}$) is
only a factor of 3 lower than observed by ASCA (Guainazzi et al. 1996)
or RXTE (Uttley et al. 1998)
in NGC~4051 ``normal'' states. This shows that a substantial contribution
to the iron line is likely {\it not} to be produced in the
immediate vicinity of the black hole, as previously suggested by
Guainazzi et al. (1996) and Uttley et al. (1998).

A prominent excess above the extrapolation of the hard X--ray best--fit
is evident below 4~keV
A detailed spectral description of the soft X--ray spectrum is
beyond the scope of this letter, and is deferred to a forthcoming 
paper. We only note that 
if a simple power--law model is added to
the model CR+BL in Table~\ref{tab1}, the soft X--ray spectral
index turns out to be ${\rm \Gamma_{soft} = 3.0\pm^{0.2}_{0.3}}$
totally inconsistent with the intrinsic
spectral index of the primary nuclear emission
(${\rm \Gamma_{primary} = 1.75\pm^{0.18}_{0.15}}$,
${\rm \chi^2 = 116.9/93}$~dof).
There is no evidence of absorption edges and/or emission lines,
the 90\% upper limit on the optical depths of the {\sc
O vii} or {\sc O viii} photoabsorption edges being 0.34 and 0.13,
respectively.
The 0.5--2~keV (0.1--2~keV) {\it observed} flux
is $\simeq 7.4 \ (12.1) \times 10^{-13}$~erg~cm$^{-2}$~s$^{-1}$, 
corresponding
to an {\it unabsorbed}
rest frame luminosity of $1.7 \ (7.0) \times 10^{40}$~erg~s$^{-1}$.

\section{The glorious past of NGC4051 (and its humble present)}

\begin{figure}
\begin{center}
\epsfig{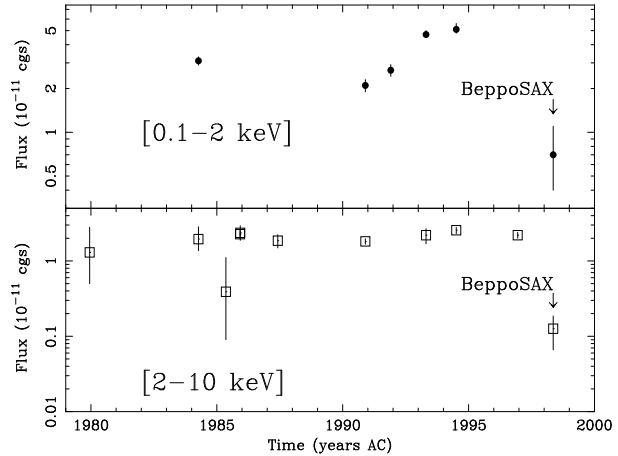}
\end{center}
\caption{NGC4051 historical 0.1--2~keV ({\it upper panel})
and 2--10~keV ({\it lower panel}) flux curves,
obtained from the high--energy X--ray spectra catalog of
Malaguti et al. 1994, except for the 1990 ROSAT/Ginga
(Walter et al. 1994), 1991 ROSAT (McHardy et al. 1995),
ASCA (1993, Mihara et al. 1994; 1994, Guainazzi et al. 1996) 
and RXTE long-look (Uttley et al. 1998)
observations.
The error bars represent the 1--$\sigma$ uncertainties on the
{\it average}
fluxes.}
\label{fig8}
\end{figure}

\begin{figure}
\begin{center}
\epsfig{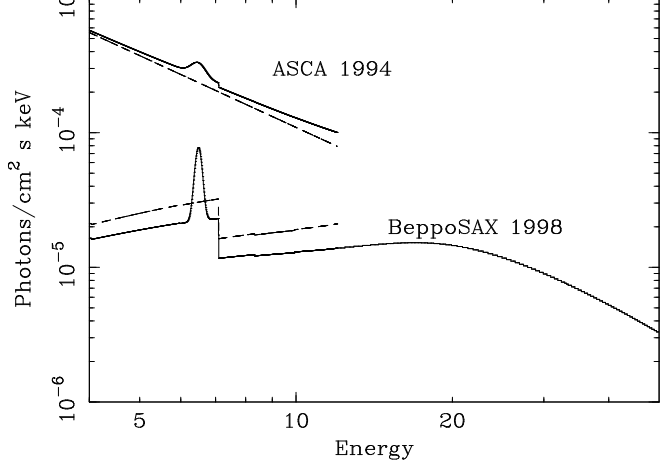}
\vspace{0.5cm}
\end{center}
\caption{Best--fit model for
the 1996 ASCA and 1998 BeppoSAX observations.
{\it Solid lines} mark the total
spectra, while {\it dashed lines} the spectra components of
the ASCA best--fit continuum
only (for which reflection from a plane--parallel
infinite slab is assumed).}
\label{fig7}
\end{figure}

NGC~4051 is well-known to exhibit large X--ray flux variability
on relatively short ($\sim 10^4$~s) timescales
(Lawrence et al. 1987;
Matsuoka et al. 1990; Guainazzi et al. 1996; Uttley et al. 1998).
BeppoSAX instead observed NGC~4051 in an ultra--dim and steady state
for the whole $\sim$~2.3 days of the observation.
The measured average 2--10~keV flux is more than one order of magnitude lower
than any historical published flux (with the exception
of a poorly constrained EXOSAT 1985 measure, see Figure~\ref{fig8}).
%-----------------------------Figure 8--------------------------------
%-----------------------------Figure 8--------------------------------
The only comparable example of such a fading is the Narrow Line
Emission Galaxy NGC~2992, which showed a flux decrease by a factor
of $\sim 20$ in 16 years (Weaver et al. 1996).
This dramatic flux decrease is associated with an equally dramatic
spectral change.
In Figure~\ref{fig7}, we show the 4--50~keV
%-----------------------------Figure 7--------------------------------
%-----------------------------Figure 7--------------------------------
best--fit models as observed by ASCA (1994, Guainazzi et al. 1996)
and Beppo-SAX.
In its ``normal'' state, the NGC4051 2--10~keV spectrum is dominated by
a continuum with typical intrinsic spectral
index of ${\rm \Gamma \simeq 1.9}$, to which a
Compton reflection component is superimposed, flattening
the spectrum above $\simeq 7-8$~keV. This BeppoSAX observation
is readily explained if the nucleus has switched off leaving the
reflection component as the only witness 
of the past brightness.
This interpretation is further supported by a contemporaneous
and even longer (8--15 May)
EUVE (0.02--0.21~keV) monitoring, which caught NGC~4051 in a much fainter and
constant state then usual (Fruscione et al., in preparation).
The implied decrease of the nuclear luminosity
(${\rm L_{nuc} < 0.14 \times 10^{41}}$~erg~s$^{-1}$)
may be explained, if the 
accretion occurs through a disc, either by a
proportional (to the luminosity) change in the accretion rate or by a 
smaller change but sufficient to trigger a transition of
the disc from a radiatively 
efficient Shakura--Sunayev disc to an inefficient advection--dominated flow.
Alternatively, it may provide support to the idea that
thermal--viscous hydrogen ionization disc
instabilities play a major role in the onset of the AGN
phenomenon, which could represent the ``outburst'' phase of accretion
from unstable discs (Burderi et. al. 1998).

The present, reflection dominated X--ray spectrum of 
NGC4051 closely resembles those 
observed in Compton--thick Seyfert 2 galaxies
(Iwasawa \& Comastri 1998; Iwasawa, Matt \& Fabian 1997;
Matt et al. 1996, 1997).
In those cases, the nucleus is likely to be completely obscured rather
than switched off, a possibility which cannot be formally
excluded by present data. It requires that a cloud of
at least ${\rm N_H > 1.4 \times 10^{26}}$~cm$^{-2}$ has obscured
the nuclear region in coincidence with the BeppoSAX pointing.
It is not easy to imagine that such an amount of absorbing
matter could be dynamically active and remain stable under
the combined action of the gravitational and radiative pressures in
the active nucleus environment. Therefore, the closest similarity occurs
with our
own Galactic Center, where the reflection spectrum and the 6.4 keV
fluorescent line from the SgrB2 
molecular cloud have been interpreted as the signature of past
activity in the now quiescent Galactic Center (Sunayev et al. 1993;
Koyama et al. 1996; Sunayev \& Churazov 1998).

The soft X-rays have undergone a decrease similar to the
hard X-rays
(see Figure~\ref{fig8}).
The steepness of
the spectrum and the lack of any ``warm absorber'' imprinting,
strongly point against a nuclear origin of the soft X--rays.
We are therefore likely to observe
a ``remnant'' of the faded
nuclear activity, normally underlying the overwhelming
primary continuum.
The intrinsic 0.1--2~keV luminosity inferred by the present
data is $\sim 7 \times 10^{40}$~erg~s$^{-1}$, a luminosity high but
not uncommon for a ``normal" galaxy. It is therefore possible that
the dimming of the nucleus has left the galactic components 
(supernovae winds, hot halos) as the bulk
of the observed soft X--rays.
Alternatively, the observed soft X-rays
could have the same origin as the extended
(spatial scale $\sim 100$~pc) emission observed by the
ROSAT HRI in NGC~4151
(Morse et al. 1995) and coinciding with the optical narrow line
emitting clouds. It was interpreted as thermal emission from
a hot (${\rm T \sim 10^7}$~K) and low density (${\rm n_e < 1}$~cm$^{-3}$)
gas, in pressure equilibrium with the Narrow Line Region clouds.
High energy and spatial resolution observations with AXAF or XMM
could likely contribute to settle this issue.

\section*{Acknowledgments}

This paper has made use of the cleaned and linearized
event files produced at the SDC.
The authors acknowledge the whole staff of the the
SDC for the
skillful management of the observation and
the prompt and careful reduction of the data.
We appreciate valuable suggestions from M.Elvis.
MG and AO acknowledge an ESA Research Fellowship.
GM and GCP acknowledge financial support from the
Agenzia Spaziale Italiana.
The BeppoSAX satellite is a joint Italian-Dutch programme.

\end{document}